\documentclass[twocolumn,preprintnumbers,prd,superscriptaddress,groupedaddress]{revtex4}
\usepackage{inputenc,psfrag,graphicx,epsfig,amsmath,amssymb,amsfonts,bbold,bm,manfnt,color,feynmp,caption,makeidx,hyperref}
\def\iscol#1#2{{\buildrel#1\parallel#2\over\longrightarrow}}
\def\beq{\begin{equation}}
\def\eeq{\end{equation}}
\def\beqa{\begin{eqnarray}}
\def\eeqa{\end{eqnarray}}

\begin{document}

\preprint{CERN-PH-TH-2015-148, CP3-15-18, Edinburgh 2015/08, LPN15-026} 

\title{\boldmath
Three-loop corrections to the soft anomalous dimension in multi-leg scattering
\unboldmath}

\author{\O{}yvind Almelid}
\affiliation{Higgs Centre for Theoretical Physics, School of Physics and Astronomy, 
The University of Edinburgh, Edinburgh EH9 3JZ, Scotland, UK}

\author{Claude Duhr}
\altaffiliation{On leave from the ``Fonds National de la Recherche Scientifique'' (FNRS), Belgium.}
\affiliation{CERN Theory Division, 1211 Geneva 23, Switzerland}
\affiliation{Center for Cosmology, Particle Physics and Phenomenology (CP3),
Universit\'{e} Catholique de Louvain, 1348 Louvain-La-Neuve, Belgium} 

\author{Einan Gardi} 
\affiliation{Higgs Centre for Theoretical Physics, School of Physics and Astronomy, 
The University of Edinburgh, Edinburgh EH9 3JZ, Scotland, UK}

\begin{abstract}

\noindent
We present the three-loop result for the soft anomalous dimension governing long-distance singularities of multi-leg gauge-theory scattering amplitudes of massless partons. 
We compute all contributing webs involving semi-infinite Wilson lines at three loops
and obtain the complete three-loop correction to the dipole formula.
We find that non-dipole corrections appear already for three coloured partons, where the correction is a constant without kinematic dependence. Kinematic dependence appears only through conformally-invariant cross ratios for four coloured partons or more, and the result can be expressed in terms of single-valued harmonic polylogarithms of weight five.
While the non-dipole three-loop term does not vanish in two-particle collinear limits, its contribution to the splitting amplitude anomalous dimension reduces to a constant, and it only depends on the colour charges of the collinear pair, thereby preserving strict collinear factorization properties.
Finally we verify that our result is consistent with expectations from the Regge limit.

\end{abstract}


\maketitle

\noindent
Infrared (long-distance) singularities are a salient feature of gauge-theory scattering amplitudes, and a 
detailed understanding of their structure and how they cancel in measurable cross sections is key to precision collider physics. For this reason,
there has been a continuous theoretical interest in the factorization and exponentiation properties of the singularities, and their use for resummation of large logarithmic corrections, starting from the analysis of the form factor in the early days~\cite{Mueller:1979ih,Collins:1980ih,Sen:1981sd,Sen:1982bt,Gatheral:1983cz,Frenkel:1984pz,Korchemsky:1985xj,Magnea:1990zb,Korchemsky:1993uz,Korchemsky:1993hr} through to many 
recent studies of multi-leg amplitudes of both 
massless~\cite{Catani:1996vz,Catasterm,Kidonakis:1998nf,Bonciani:2003nt,Dokshitzer:2005ig,Aybat:2006mz,Dixon:2008gr,Dixon:2009gx,Gardi:2009qi,Becher:2009cu,Gardi:2009zv,Dixon:2009ur,Bret:2011xm,Korchemsky:1993hr,Caron-Huot:2013fea,Erdogan:2014gha,Ahrens:2012qz,Gehrmann:2010ue,Naculich:2013xa} and massive partons~\cite{Kidonakis:2009ev,Mitov:2009sv,Becher:2009kw,Beneke:2009rj,Czakon:2009zw,Ferroglia:2009ep,Ferroglia:2009ii,Chiu:2009mg,Mitov:2010xw,Gardi:2013saa,Henn:2013tua}
at the multi-loop level, and the formulation of the non-Abelian exponentiation theorem in the multi-leg case~\cite{Gardi:2010rn,Mitov:2010rp,Gardi:2011wa,Gardi:2011yz,Gardi:2013ita}.

The focus of this paper will be the infrared (IR) structure of a scattering amplitude for $n$ massless partons. More precisely,
if the external legs have momenta $p_i$, $i=1..n$, with $p_i^2=0$, long distance singularities (both soft and collinear) can be factorized as follows 
\begin{equation}
{\cal M}_n\left(\left\{p_i\right\}, \alpha_s \right) =
Z_n \left(\left\{p_i\right\}, \mu, \alpha_s \right) 
{\cal H}_n \left(\left\{p_i\right\}, \mu, \alpha_s \right)\,, 
\end{equation}
where $\mu$ is a factorization scale, $\alpha_s\equiv \alpha_s(\mu^2)$ is the renormalised $D$-dimensional running coupling, ${\cal H}_n$ is a {finite} hard scattering function, and $Z_n$ is an operator in colour space that collects all IR singularities in the form of poles in the dimensional regularization parameter $\epsilon = (4-D)/2$. The IR singularities contained in $Z_n$ have their origin in loop momenta becoming either soft or collinear to any of the scattered partons (see e.g. Ref.~\cite{Sterman:1995fz}). Collinear singularities depend on the spin and momentum of that particle, and decouple from the rest of the process. In contrast, soft (non-collinear) singularities are independent of the spin, but they depend on the relative directions of motion and the colour degrees of freedom of the scattered particles. Hence, soft singularities are sensitive to the colour flow in the entire process,  and their structure is a priori rather complex. Nevertheless, they are significantly simpler than finite contributions to the amplitude. They can be computed by considering correlators of products of Wilson-line operators emanating from the hard interaction, following the classical trajectory of the scattered particles and carrying the same colour charge. 

{
\begin{figure*}[!t]
\begin{center}
\scalebox{.25}{\includegraphics{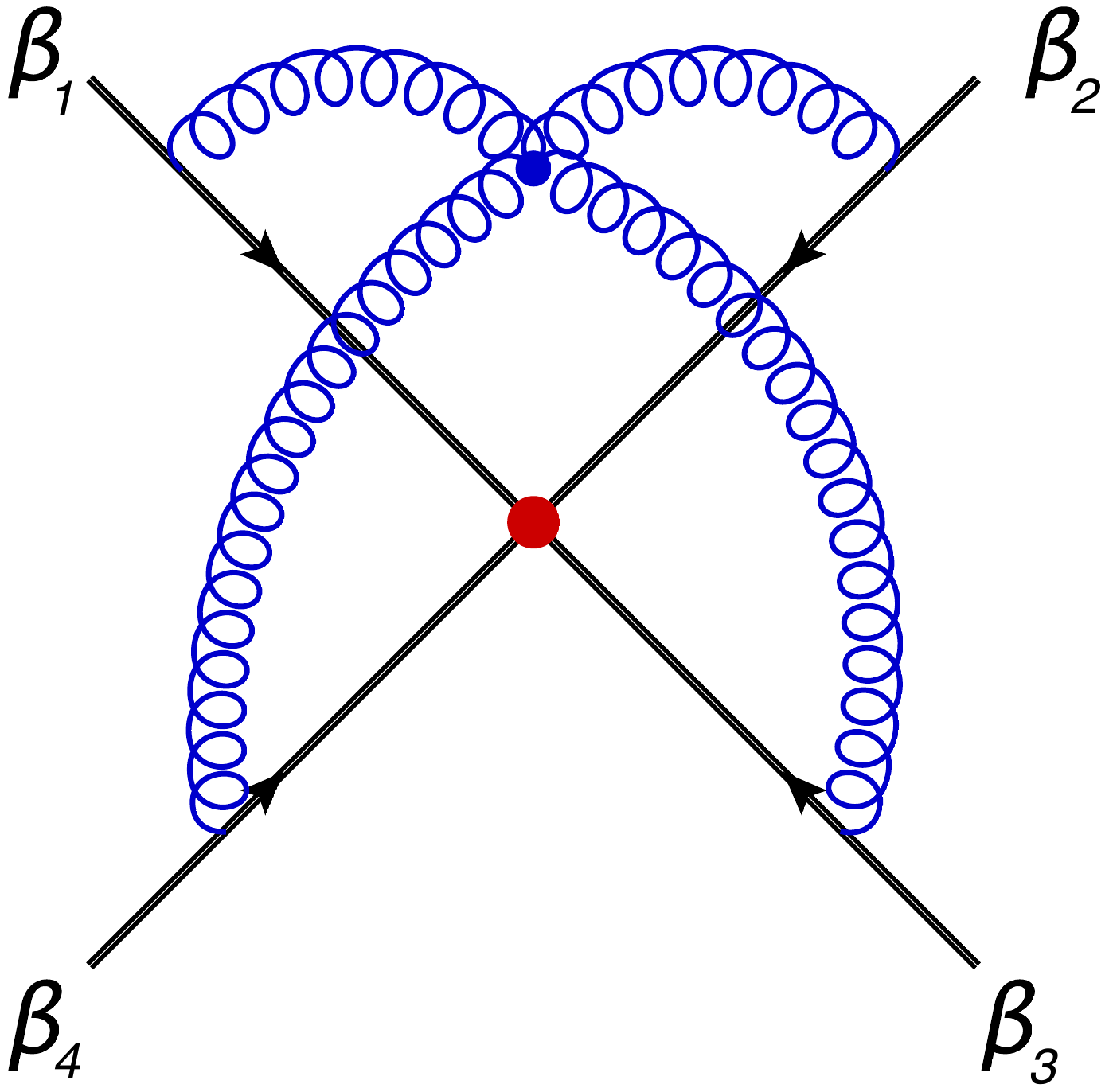}}\hspace*{10pt}
\scalebox{.25}{\includegraphics{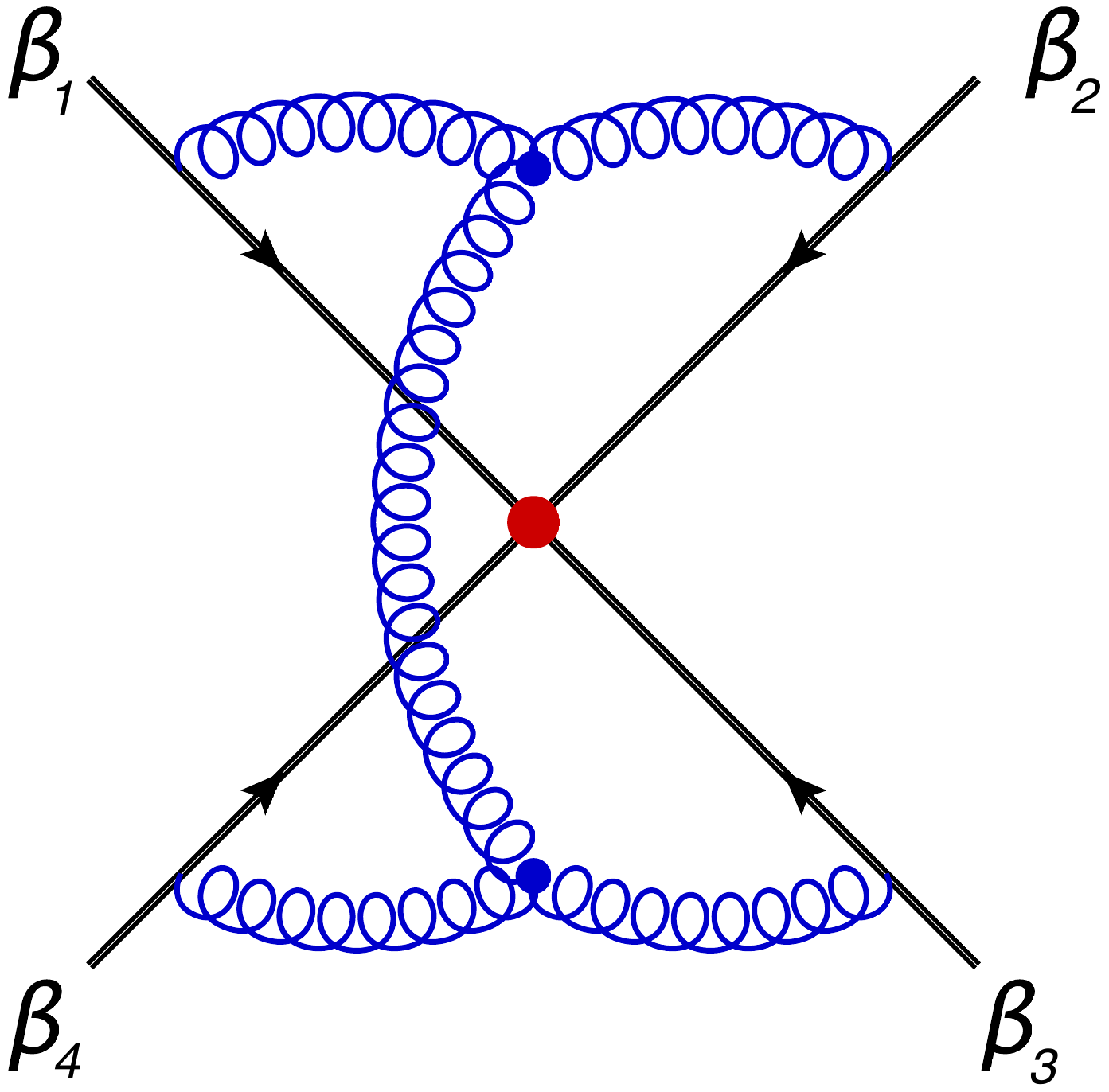}}\hspace*{10pt}
\scalebox{.25}{\includegraphics{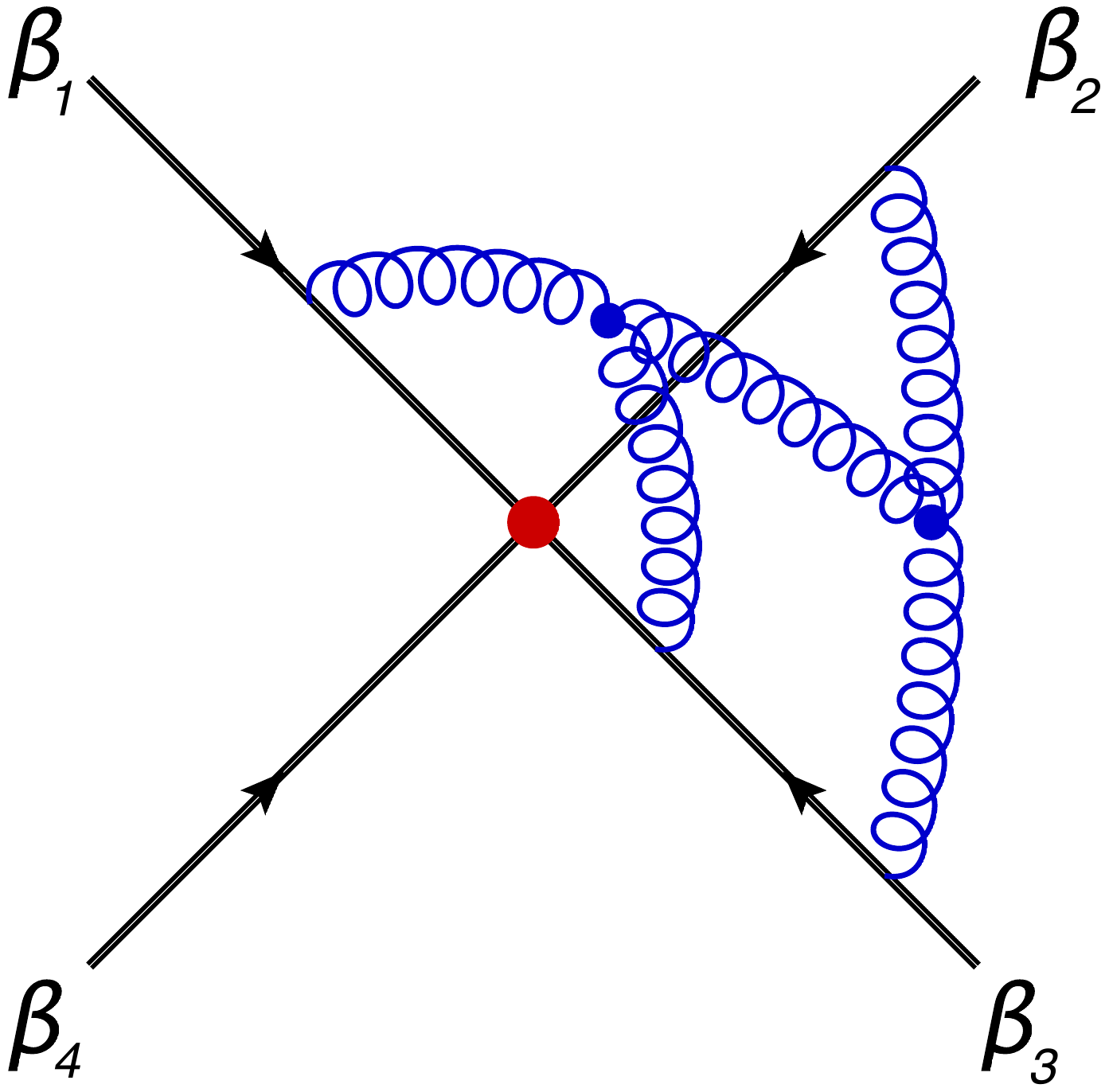}}\hspace*{10pt}
\scalebox{.25}{\includegraphics{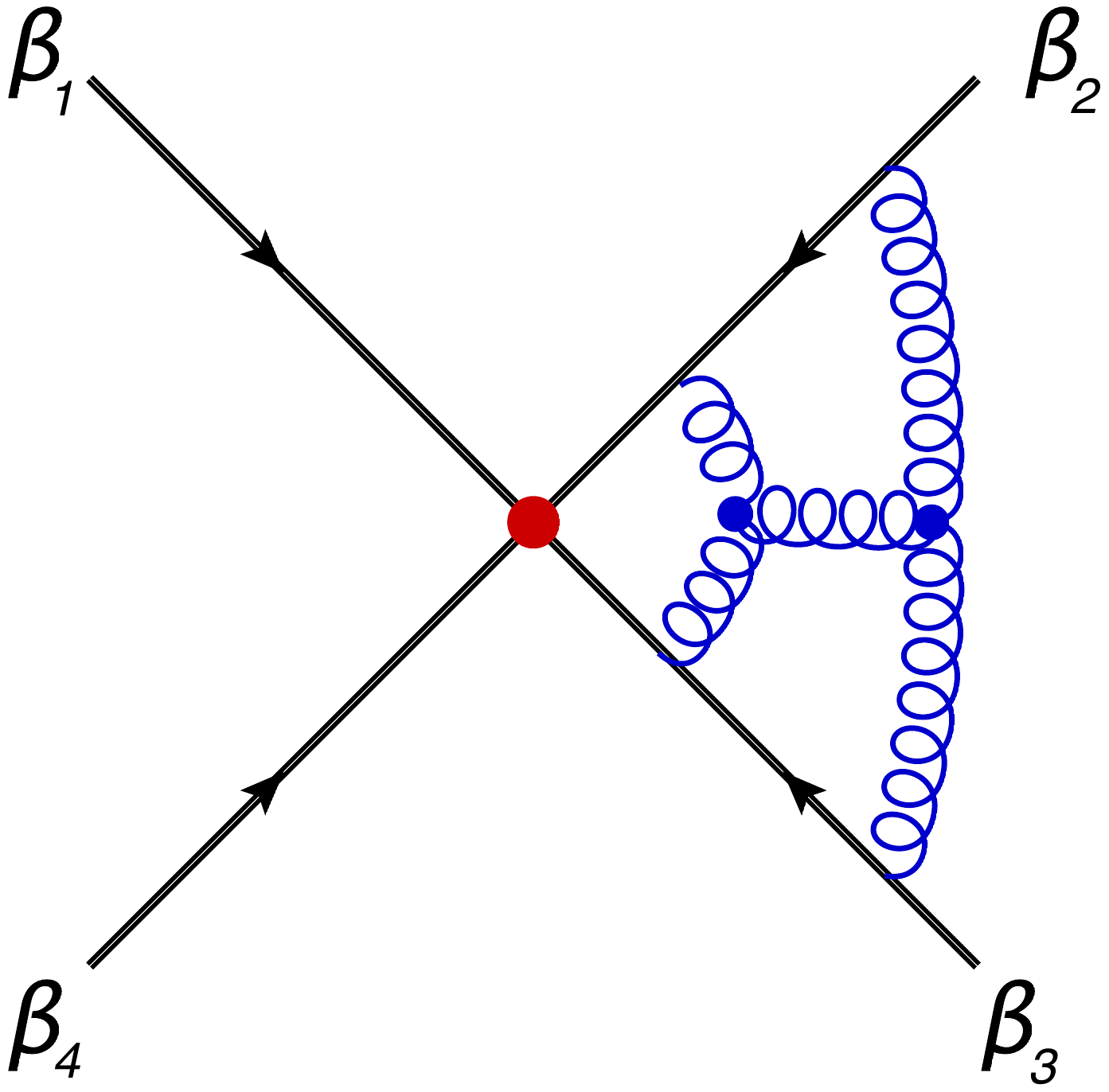}}
\caption{Representative 3-loop connected webs contributing to the soft anomalous dimension with 4 coloured lines.}
\label{4lines_connected}
\end{center}
\end{figure*}}

Specifically, $Z_n$ can be obtained as a solution of a renormalization-group equation as
\begin{equation}\begin{split}
\label{sumodipoles}
\! Z_n&   = {\rm P}
  \exp \Bigg\{\! -\frac12 \int_0^{\mu^2}\! \frac{d \lambda^2}{\lambda^2} \Gamma_n\left(\left\{p_i\right\}, \lambda,\alpha_s(\lambda^2)\right)\! \Bigg\}\,,
\end{split}\end{equation}
where $\Gamma_n$ is the so-called \emph{soft anomalous dimension matrix} for multi-leg scattering, and ${\rm P}$ stands for path-ordering of the matrices according to the order of scales $\lambda$.  We stress that $\Gamma_n$ itself is finite, and IR singularities are generated in Eq.~\eqref{sumodipoles} owing to the fact that $\Gamma_n$ depends on the $D$-dimensional coupling,
which is integrated over the scale down to zero momentum. The functional form of $\Gamma_n$ is highly constrained, and 
owing to factorization and the rescaling symmetry of the Wilson line velocities~\cite{Gardi:2009qi,Becher:2009cu,Gardi:2009zv}, through three loops it must take the form
\begin{align}
\label{Gamma}
\begin{split}
\Gamma_n
\left(\left\{p_i\right\}, \lambda\right) 
=
\Gamma_{n}^{\rm dip.}
\left(\left\{p_i\right\}, \lambda\right) 
+\Delta_n\left(\left\{\rho_{ijkl}\right\}\right)\,,
\end{split}
\end{align}
with
\beqa
\label{dipole_formula}
\Gamma_{n}^{\rm dip.} \left(\left\{p_i\right\}, \lambda\right) = & -&\frac{1}{2} \widehat{\gamma}_K \left( \alpha_s \right) 
  \sum_{i<j}  \log \left(\frac{-s_{ij}}{\lambda^2} \right)
  {\bf T}_i \cdot {\bf T}_j \nonumber \\ &  +& \sum_{i = 1}^n
  \gamma_{J_i} \left(\alpha_s\right) \,,
\eeqa
where $- s_{ij} = 2 \left\vert p_i \cdot p_j \right \vert e^{ - {\rm i}
\pi \lambda_{ij}}$, with $\lambda_{i j} = 1$ if partons
 $i$ and $j$ 
both belong to either the initial or the final state and $\lambda_{i j} = 0$ 
otherwise; ${\bf T}_i$ are colour generators in the representation of 
parton $i$, acting on the colour indices of the amplitude as described 
in Ref.~\cite{Catani:1996vz}; $\widehat{\gamma}_K (\alpha_s)$ is the universal cusp anomalous dimension~\cite{Korchemsky:1985xj,Grozin:2014hna,Moch:2004pa}, with the quadratic Casimir of the appropriate representation scaled out (Casimir scaling of the cusp anomalous dimension holds through three loops~\cite{Moch:2004pa}; it may be broken by 
quartic Casimirs starting at four loops); $\gamma_{J_i}$ 
are the anomalous dimensions of the fields associated with external 
particles, which govern hard collinear singularities, currently known up to three loops~\cite{Moch:2005tm,Gehrmann:2010ue}. Equation~\eqref{dipole_formula} is known as the \emph{dipole formula}, and captures the entirety of the soft anomalous dimension matrix up to two loops. 
According to the non-Abelian exponentiation theorem~\cite{Gardi:2013ita} the colour factors in $\Delta_n$ must all correspond to connected graphs as shown in Fig.~\ref{4lines_connected}.
Tripole corrections correlating three partons, with colour factors 
of the form ${\rm i}f^{abc} {\rm \bf T}_i^a{\rm \bf T}_j^b{\rm \bf T}_k^c$,
which could appear starting from two loops, are not present in the soft anomalous dimension at any order because the corresponding kinematic dependence on the three momenta is bound to violate the rescaling symmetry constraints~\cite{Gardi:2009qi,Becher:2009cu,Gardi:2009zv}.
While a constant correction proportional to ${\rm i}f^{abc} {\rm \bf T}_i^a{\rm \bf T}_j^b{\rm \bf T}_k^c$ is excluded by Bose symmetry, kinematic-independent corrections involving three lines of the form  $f^{abe}f^{cde}\left\{{\rm \bf T}_i^a,  {\rm \bf T}_i^d\right\}   {\rm \bf T}_j^b {\rm \bf T}_k^c$ 
(last two diagrams in 
Fig.~\ref{4lines_connected}) 
are admissible and we will see that they do indeed appear.   
The first admissible corrections involving kinematic dependence in Eq.~\eqref{Gamma} are then quadrupoles, because four momenta can form conformally-invariant cross ratios, 
\beq\label{eq:CICR}
\rho_{ijkl}\equiv\frac{(-s_{ij})(-s_{kl})}{(-s_{ik})(-s_{jl})}\,,
\eeq
which are invariant under a rescaling of any of the momenta. Since diagrams with four colour generators contribute for the first time at three loops, this is the first order at which contributions to $\Delta_n$ in Eq.~\eqref{Gamma} may appear, 
\begin{equation}
\Delta_n\left(\left\{\rho_{ijkl}\right\}\right) = \sum_{\ell=3}^\infty\left(\frac{\alpha_s}{4\pi}\right)^\ell \Delta_n^{(\ell)}\left(\left\{\rho_{ijkl}\right\}\right)\,.
\end{equation}
Three-loop graphs can connect at most four lines, and so the general form of the three-loop correction is completely determined by the four-parton case and can be written as
\begin{align}\label{eq:expected_Delta}
 &\Delta_n^{(3)}\left(\left\{\rho_{ijkl}\right\}\right) = 16\,f_{abe}f_{cde} 
\Big\{\\ 
\nonumber
& 
{\sum_{1\leq i<j<k<l\leq n}}
\Big[
 {\rm \bf T}_i^a  {\rm \bf T}_j^b   {\rm \bf T}_k^c {\rm \bf T}_l^d   \, {\cal F}(\rho_{ikjl},\rho_{iljk}) 
 \\[-8pt]
 \nonumber
&\phantom{{\sum_{1\leq i<j<k<l\leq n}}}\hspace*{-8pt}+{\rm \bf T}_i^a  {\rm \bf T}_k^b {\rm \bf T}_j^c   {\rm \bf T}_l^d    
\, {\cal F}(\rho_{ijkl},\rho_{ilkj}) \\[-8pt]
\nonumber 
&\phantom{{\sum_{1\leq i<j<k<l\leq n}}}\hspace*{-8pt}+ {\rm \bf T}_i^a   {\rm \bf T}_l^b  {\rm \bf T}_j^c    {\rm \bf T}_k^d 
\, {\cal F}(\rho_{ijlk},\rho_{iklj}) \Big]
\\[-14pt]
\nonumber&-C\, \sum_{i=1}^n\sum_{\substack{{1\leq j<k\leq n}\\ j,k\neq i}}\left\{{\rm \bf T}_i^a,  {\rm \bf T}_i^d\right\}   {\rm \bf T}_j^b {\rm \bf T}_k^c   
\Big\}\,,
\end{align}
where $C$ is a constant and ${\cal F}$ is a function of two conformally-invariant cross ratios.  Both $C$ and $\mathcal{F}$ are independent of the colour degrees of freedom. 
Moreover, Eq. (\ref{eq:expected_Delta}) is the most general three-loop ansatz consistent with Bose and rescaling symmetry, so $C$ and $\mathcal{F}$ are independent of the number of legs $n$.
Note that the terms in this sum are not all independent, because of the antisymmetry of the structure constants and the Jacobi identity. $\Delta_n^{(3)}$ is independent of the details of the underlying theory and completely determined by soft gluon interactions. In particular, this implies that $\Delta_n^{(3)}$ is the same in QCD and in $\mathcal{N}=4$ Super Yang-Mills, and it is therefore expected to be a pure polylogarithmic function of weight five. Its functional form has been constrained by considering collinear limits and the Regge limit ~\cite{Gardi:2009qi,Dixon:2009gx,Becher:2009cu,Dixon:2009ur,Gardi:2009zv,Bret:2011xm,Ahrens:2012qz,Naculich:2013xa,Caron-Huot:2013fea}, but it has so far remained unclear whether three-loop corrections to the dipole formula are present. 
The purpose of the present paper is to compute $\Delta_n^{(3)}$. We will present its complete functional form, hence determining soft singularities of any massless multi-leg amplitude at three loops. Since $C$ and $\mathcal{F}$ can be extracted from $\Delta_4^{(3)}$, we restrict our computation to the case $n=4$. Before presenting the final result, we give a brief summary of the computation. A complete account of the computation will be presented in a forthcoming publication~\cite{longpaper}.

\begin{figure*}[t]
\begin{center}
\scalebox{.25}{\includegraphics{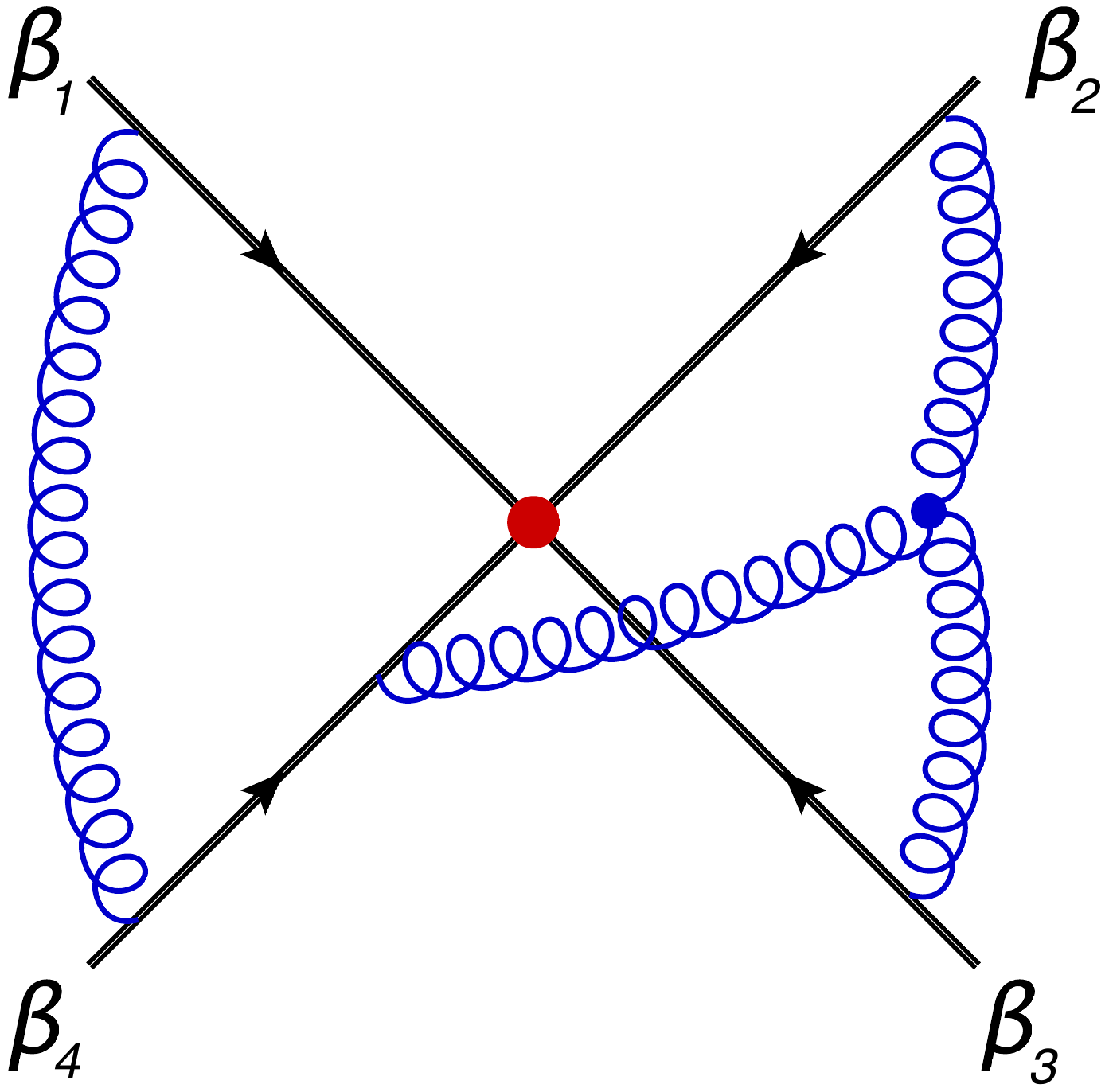}}
\hskip 2cm
\scalebox{.25}{\includegraphics{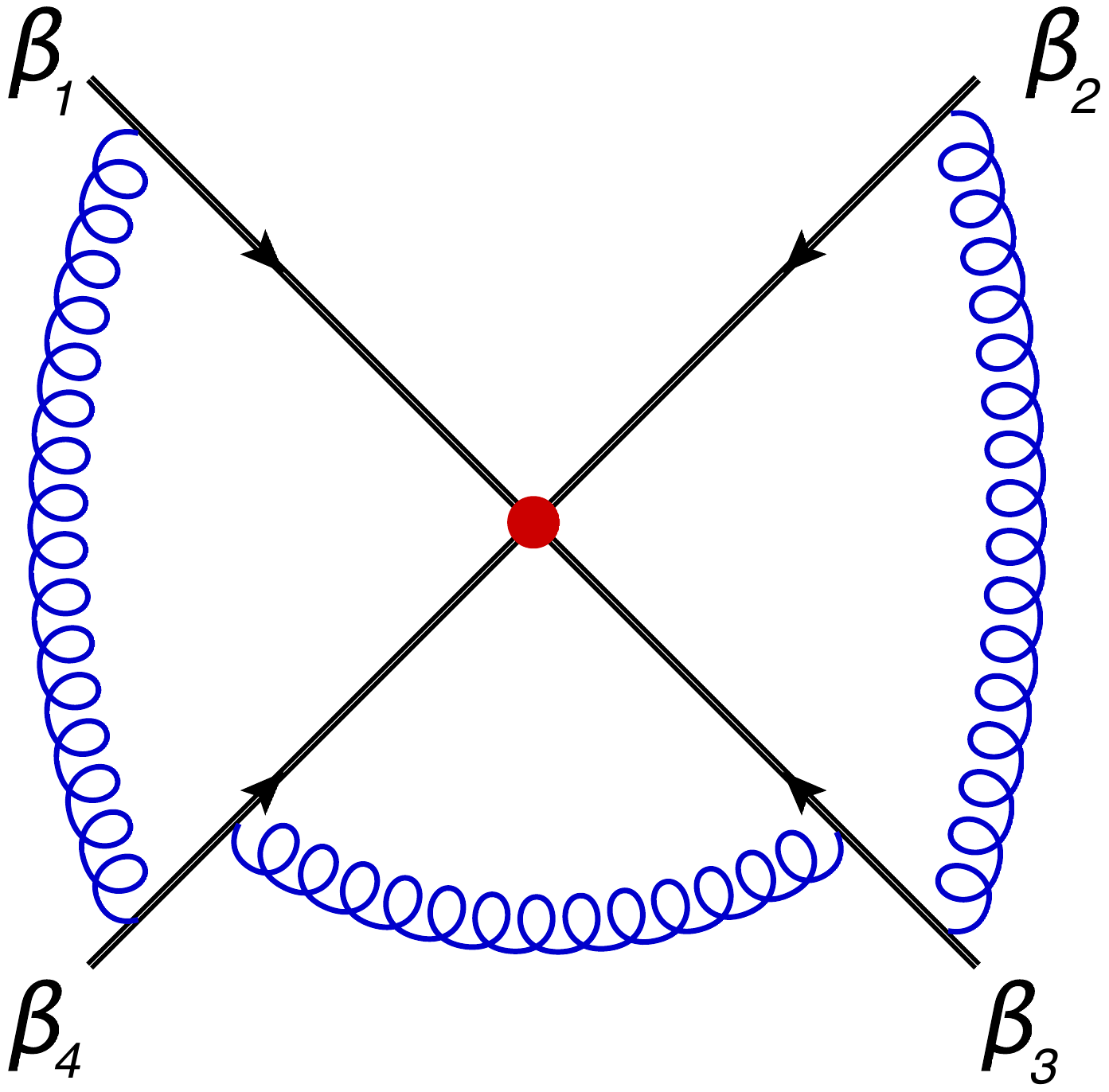}}
\hskip 2cm
\scalebox{.25}{\includegraphics{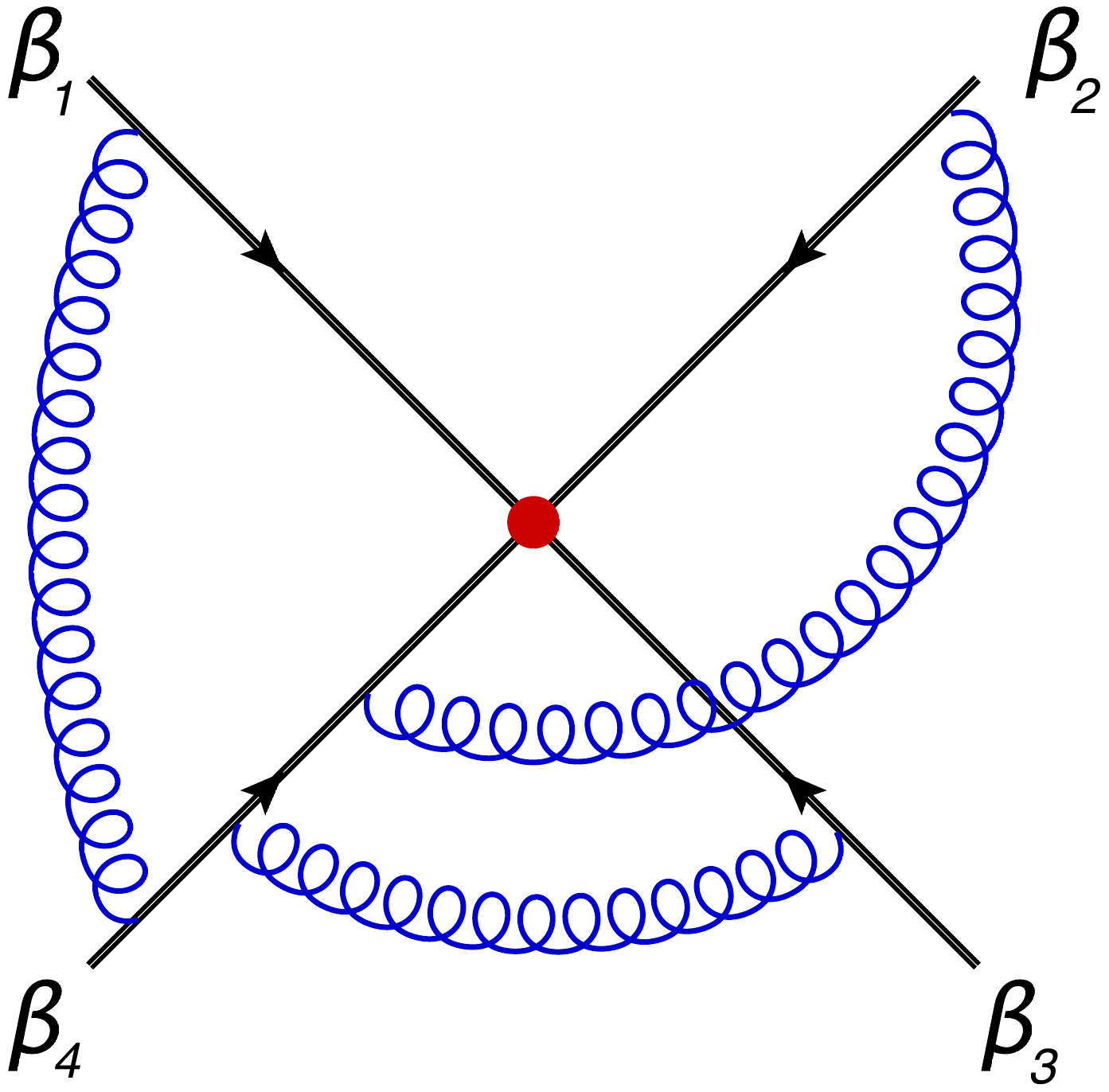}}
\caption{Representative non-connected three-loop diagrams of webs which contribute to the quadrupole term $\Delta_4^{(3)}$.}
\label{4lines_non-connected}
\end{center}
\end{figure*}

We set up the calculation of the soft anomalous dimension through the renormalization of a product of semi-infinite Wilson lines with four-velocities $\beta_k$, with $\beta_k^2\neq 0$. By considering non-lighlike lines we avoid collinear singularities, and obtain kinematic dependence via cusp angles $\gamma_{ij}\equiv 2\beta_i\cdot\beta_j/\sqrt{\beta_i^2\beta_j^2}$.
We eventually extract $\Delta_n^{(3)}$ for massless scattering by considering the asymptotic lightlike limit $\beta_k^2\to 0$, where the kinematic dependence reduced to conformally-invariant cross ratios as in Eq.~\eqref{eq:CICR}.

 In organising the calculation we use of the non-Abelian exponentiation theorem and we only compute \emph{webs}. A web can be either an individual connected diagram, as in Fig.~\ref{4lines_connected}, or a set of non-connected diagrams which are related by permuting the order of gluon attachments to the Wilson lines~\cite{Gardi:2010rn,Mitov:2010rp,Gardi:2011wa,Gardi:2011yz,Gardi:2013ita}; representative diagrams from such webs are shown in Fig.~\ref{4lines_non-connected}. In either of these cases, the contribution to $\Delta_4^{(3)}$ is associated with fully connected colour factors, the classification of which was done in Ref.~\cite{Gardi:2013ita}. The sum of all two-line diagrams can be written in the form
\begin{align*}
{\small 
G_2(1,2)=\text{dipole}-f^{abe}f^{cde} 
\left\{ {\rm \bf T}_1^a, {\rm \bf T}_2^d\right\} \left\{{\rm \bf T}_1^b, {\rm \bf T}_2^c\right\} \, H_2(1,2)\,,}
\end{align*}   
where `dipole' stands for a term with a colour factor proportional to ${\bf T}_1\cdot {\bf T}_2$, which contributes to $\Gamma_{n}^{\rm dip.}$. The component involving four generators via anti-commutators is relevant for the calculation of $\Delta_n$; its kinematic dependence is contained in $H_2(1,2)=H_2(2,1)$. Similarly, the sum of all three-line diagrams takes the form
\begin{align*}
G_3(1,2,3)= f^{abe}f^{cde}\!\!\!\!\!\!\!\!\sum_{\substack{(i,j,k)\in (1,2,3)\\j<k}} \!\!\!\!\!\!\!\left\{ {\rm \bf T}_i^a, {\rm \bf T}_i^d\right\} {\rm \bf T}_j^b {\rm \bf T}_k^c \, H_3(i,j,k) \,,
\end{align*}  
with $H_3(i,j,k) = H_3(i,k,j)$. The kinematic functions $H_2$ and $H_3$ are necessarily polynomials in $\log(-\gamma_{ij})$ for lightlike kinematics where $\gamma_{ij}\to -\infty$. Note that in the expression for $G_3(1,2,3)$ we omitted the tripole term, proportional to $f^{abc} {\rm \bf T}_1^a {\rm \bf T}_2^b {\rm \bf T}_3^c$, which vanishes in this kinematic limit.
Finally, three-loop webs connecting four lines can be cast into the form
\begin{align}
\nonumber
&G_4(1,2,3,4) = {\rm \bf T}_1^a {\rm \bf T}_2^b {\rm \bf T}_3^c {\rm \bf T}_4^d \Big[f^{abe}f^{cde} H_4(1,2,3,4)\,+
\\&  f^{ace}f^{bde} H_4(1,3,2,4) + f^{ade}f^{bce} H_4(1,4,2,3)\Big]\,,
\label{G4}
\end{align}
where kinematic function $H_4$ satisfies $H_4(1,2,3,4)= -H_4(2,1,3,4)= H_4(3,4,1,2)$;
this function depends on logarithms of cups angles as well as on non-trivial functions of cross ratios as in Eq.~\eqref{eq:CICR}.

Another important element in organising the calculation is colour conservation. The anomalous dimension $\Gamma_n$ is an operator in colour space that acts on the hard amplitude, which is a colour singlet and must therefore satisfy~\cite{Catasterm}
\begin{align}
\begin{split}
\left(\sum_{i=1}^n{\rm \bf T}_i^a\right)\mathcal{H}_n = 0\,.
\end{split}
\end{align}
Hence, when computing $\Delta_4^{(3)}$ one may systematically eliminate ${\rm \bf T}_4$ in favour of ${\rm \bf T}_i$, $1\le i\le3$, thereby reducing all four-line colour factors such as $f^{abe}f^{cde} {\rm \bf T}_i^a  {\rm \bf T}_j^b {\rm \bf T}_k^c {\rm \bf T}_l^d$ to three-line ones such as $f^{abe}f^{cde} \left\{{\rm \bf T}_i^a,  {\rm \bf T}_i^d\right\}   {\rm \bf T}_j^b {\rm \bf T}_k^c$. As a consequence, colour conservation relates sets of diagrams that connect a different numbers of Wilson lines.
Summing over all subsets of two and three lines out of four and using colour conservation, we find that the function $\mathcal{F}$ and the constant $C$ can be expressed in terms of the kinematic functions $H_n$ as follows:
\begin{align}
\label{calF_in_terms_of_H}
&\!\!\mathcal{F}(\rho_{ijkl},\rho_{ilkj})= H_4(i,j,k,l)
-\frac23 \Big[\overline{H}_3(i,j,k)
\\&\!\!-\overline{H}_3(i,j,l)
-\overline{H}_3(j,i,k)+\overline{H}_3(j,i,l)+\overline{H}_3[k,\{i,l\}]
\nonumber
\\&\!\!-\overline{H}_3(k,j,l)
-\overline{H}_3(l,i,k)
+\overline{H}_3(l,j,k)\Big]\,,\nonumber\\
\label{C_in_terms_of_H3}
&\!\!C= \frac13\Big[\overline{H}_3(i,j,k) +\overline{H}_3(j,k,i) +\overline{H}_3(k,j,i)\Big]\,.
\end{align}   
where {\small $\overline{H}_3(i,j,k)= H_3(i,j,k) +H_2(i,j) +H_2(i,k)$}. 
The above equations put strong constraints on the kinematic functions $H_n$: the function $\mathcal{F}$ depends on conformal cross ratios, while the functions $H_n$ on the right-hand side of Eq.~(\ref{calF_in_terms_of_H}) depend of logarithms of cusp angles; these must therefore conspire to combine into logarithms of conformal cross ratios. In addition, $C$ is a constant, and so the kinematic dependence of the functions $H_n$ must cancel in the sum in Eq.~\eqref{C_in_terms_of_H3}. Our computation satisfies all these constraints, which provides a strong check of the result.

The calculation of the individual graphs is rather lengthy, and we will only describe the main steps, deferring a detailed exposition to a dedicated publication~\cite{longpaper}. We set up the calculation in configuration space, with four non-lightlike Wilson lines with four-velocities $\beta_k$. The position of the three- and four-gluon vertices off the Wilson lines are integrated over in $D=4-2\epsilon$ dimensions. Following Ref.~\cite{Gardi:2011yz,Gardi:2013saa}, we introduce an infrared regulator which suppresses exponentially contributions far along the Wilson lines. This is necessary to capture the ultraviolet singularity associated with the vertex where the Wilson lines meet. Upon performing the integral over the overall scale, we extract an overall $1/\epsilon$ ultraviolet pole, and the contribution to the soft anomalous dimension is the coefficient of that pole, which is finite for each of the diagrams in Fig.~\ref{4lines_connected} (they have no subdivergences) and can be evaluated in $D=4$ dimensions.
Next, we observe that the integrals  
over
the positions of the three- and four-gluon vertices give rise to one- and two-loop off-shell four-point functions, for which we  derive a multifold Mellin-Barnes (MB) representation. After integration over the position of the gluon emission vertex along the Wilson lines, we obtain a MB representation of each of the connected graphs for the general non-lightlike case, depending on the velocities through the cusp angles $\gamma_{ij}$. In order to proceed, we use standard techniques~\cite{Czakon:2005rk} to perform a simultaneous asymptotic expansion for $\gamma_{ij}\to -\infty$ corresponding to the lightlike limit, where we neglect any term suppressed by powers of $1/\gamma_{ij}$. After this procedure, we obtain a collection of lower-dimensional MB integrals. The remaining MB integrals are then converted into parametric integrals using the methods of Ref.~\cite{Anastasiou:2013srw}, which can be performed using modern integration techniques~\cite{Goncharov.A.B.:2009tja}. The sum over all connected graphs is expressible as a linear combination of products of logarithms of cusp angles $\gamma_{ij}$ and single-valued harmonic polylogarithms~\cite{BrownSVHPLs,Remiddi:1999ew} with arguments $z_{ijkl}$ and $\bar{z}_{ijkl}$, related to the conformally-invariant cross ratios~\eqref{eq:CICR} by
\beq
\label{zbarzdef}
z_{ijkl}\,\bar{z}_{ijkl} = \rho_{ijkl} {\rm~~and~~} (1-z_{ijkl})\,(1-\bar{z}_{ijkl})=\rho_{ilkj}\,.
\eeq
We observe that individual graphs are not pure functions, but they involve pure functions of weight five multiplied by rational functions in $z_{ijkl}$ and~$\bar{z}_{ijkl}$. These rational functions cancel in the sum over all connected graphs, leaving behind a pure function of weight five, in agreement with the expectation that scattering amplitudes in $\mathcal{N}=4$ Super Yang-Mills have a uniform maximal weight. Moreover, mixed weight terms do appear in two-line and three-line webs, but cancel out in the sum.

Adding up all the contributions, we find the following results for the function $\mathcal{F}$ and the constant $C$,
\begin{equation}\begin{split}
\mathcal{F}(\rho_{ijkl},\rho_{ilkj}) &\, = F(1-z_{ijkl}) - F(z_{ijkl})\,,\\
C &\,= \zeta_5 + 2\zeta_2\,\zeta_3\,,
\end{split}\end{equation}
with
\begin{equation}
F(z)\, = \mathcal{L}_{10101}(z) +2\,\zeta_2\,\left[\mathcal{L}_{001}(z)+\mathcal{L}_{100}(z)\right]\,,
\label{eq:F_definition}
\end{equation}
where $\mathcal{L}_{w}(z)$ are Brown's single-valued harmonic polylogarithms (SVHPLs)~\cite{BrownSVHPLs} (see also Ref.~\cite{Dixon:2012yy}). Note that we kept implicit the dependence of these functions on $\bar{z}$.  SVHPLs can be expressed in terms of ordinary harmonic polylogarithms (HPLs)~\cite{Remiddi:1999ew} in $z$ and $\bar{z}$. The result for $F$ in terms of HPLs is attached in computer-readable format to this paper.

Let us now briefly discuss the main features of the final result. First, we note that
while $F(z)$ is defined everywhere in the physical parameter space, it is only single-valued in the part of the Euclidean region (the region where all invariants are negative) where $z$ and $\bar{z}$ are complex conjugate to each other. Single-valuedness ensures that $\Delta_4^{(3)}$ has the correct branch cut structure of a physical scattering amplitude~\cite{Gaiotto:2011dt,Dixon:2012yy}: it is possible to analytically continue the function to the entire Euclidean region while the function remains real throughout~\cite{Chavez:2012kn}. Next note that if one considers $F(z)$ as a function of two independent variables $z$ and $\bar{z}$ (not a complex conjugate pair) this function has branch points for $z$ and $\bar{z}$ at $0$, $1$ and $\infty$. Crossing symmetry, i.e., crossing some momenta from the final to the initial state, is realized in a very simple way by taking monodromies around these points.

Next, let us discuss the symmetries of the final answer for the three-loop corrections to the soft anomalous dimension. In the four-line case, 
Bose symmetry is realised on the cross ratios by the action of the group $S_3$ which keeps the momentum $p_1$ fixed and permutes the remaining three momenta. This group  naturally acts on the space of SVHPLs by change of arguments generated by the transformations $(z,\bar{z})\mapsto (1-\bar{z}, 1-z)$ and $(z,\bar{z})\mapsto (1/\bar{z}, 1/z)$, with $z\equiv z_{1234}$. We note that geometrically this symmetry simply acts by exchanging the three singularities at $z\in\{0,1,\infty\}$. Moreover, it is known that the space of all HPLs, and hence also SVHPLs, is closed under the action of this $S_3$, giving rise to functional equations among HPLs, i.e., relations among HPLs with different arguments. As a consequence, it is possible to express all the terms in Eq.~\eqref{eq:F_definition} in terms of SVHPLs with argument $z$.

Besides the action of this group $S_3$, there is a second symmetry group $\mathbb{Z}_2$ acting on the space of SVHPLs. Indeed, the definition of $(z,\bar{z})$ in Eq.~(\ref{zbarzdef}) is invariant under the exchange $z\leftrightarrow \bar{z}$, and hence the function $F(z)$ must be invariant under this transformation, i.e., $F(z)$ must be an even function: $F(\bar{z}) = F(z)$. This symmetry is realised on the space of SVHPLs by the operation of reversal of words, namely, if $w$ is a word made out of 0's and 1's, and $\widetilde{w}$ the reversed word, then we have $\mathcal{L}_w(\bar{z}) = \mathcal{L}_{\widetilde{w}}(z) +\ldots$, where the dots indicate terms proportional to  multiple zeta values. Even functions then correspond to `palindromic' words (possibly up to multiple zeta values), and indeed we see that Eq.~\eqref{eq:F_definition} is `palindromic'.

Let us now comment on the momentum conserving limit of $\Delta_4^{(3)}$, which is of particular interest because it corresponds to two-to-two  massless  scattering. In this limit we have $\bar{z} = z = s_{12}/s_{13} = -s/(s+t)$. It follows that for two-to-two massless  scattering $F(z)$ can be expressed entirely in terms of HPLs with indices $0$ and $-1$ depending on $s/t$, in agreement with known results for on-shell three-loop four-point integrals~\cite{Smirnov:2003vi,Bern:2005iz,Henn:2013tua}.
Furthermore, specialising to the Regge limit and  expanding Eq.~\eqref{eq:F_definition} at large $s/(-t)$ \footnote{Taking the Regge limit requires analytic continuation to the physical region of $2\to 2$ scattering, to be discussed in detail in~\cite{longpaper}.} we find no $\alpha_s^3\, \ln^p\left({s}/({-t})\right)$ for any~$p>0$: $\Delta_4^{(3)}$ simply tends to a constant in this limit. This is entirely consistent with the behaviour of a two-to-two scattering amplitude in the Regge limit~\cite{Bret:2011xm,Caron-Huot:2013fea,Caron-Huot:TBA}; indeed, the dipole formula alone is consistent with predictions from the Regge limit through next-to-next-to-leading logarithms at ${\cal O}(\alpha_s^3)$~\cite{Caron-Huot:TBA}.

Finally, let us comment on the behaviour of $\Delta_n^{(3)}$ in the limit where two final-state partons become collinear. 
A well-known property of an $n$-parton scattering amplitude is that the limit where any two coloured partons become collinear can be related to an $(n-1)$-parton amplitude: 
\begin{align} 
\label{Sp_M}
\!\!{\cal M}_n \left(p_1,p_2,\{p_j\}\right)\,\,
 \iscol{1}{2} \,\,\,{\bf Sp}(p_1,p_2) \,
{\cal M}_{n-1} \left( P, \{p_j\}\right)\, ,
\end{align}
where $P=p_1+p_2$,  and $p_j$ are the momenta of the $(n-2)$ non-collinear partons. The splitting amplitude ${\bf Sp}\left( p_1, p_2 \right)$ is an operator in colour space which captures the singular terms for $P^2\to 0$. 
All elements in Eq.~\eqref{Sp_M} have infrared singularities, and these must clearly be related. Furthermore, ${\bf Sp}$ is expected to only depend on the quantum numbers of the collinear pair~\cite{SplittingAmplitudeRefs} to all orders in perturbation theory, and hence also its soft anomalous dimension,
\begin{equation}
\Gamma_{\rm \bf Sp}=\left.(\Gamma_n-\Gamma_{n-1})\right\vert_{1\parallel 2} = \Gamma^{\rm dip.}_{\rm {\bf Sp}} +\Delta_{\rm {\bf Sp}}\,,
\end{equation}
must be independent of the momenta and colour degrees of freedom of the rest of the process. 
This property is automatically satisfied for the dipole formula, but it is highly non-trivial for it to persist when quadrupole corrections are present, 
as these might introduce correlations between the collinear pair and the rest of the process.
In Refs.~\cite{Becher:2009cu,Dixon:2009ur} this property was used to constrain $\Delta_n$, but this was done under the assumption that $C$ in Eq.~\eqref{eq:expected_Delta} vanishes. Given our result for $\Delta_n^{(3)}$, we may now compute the non-dipole correction to the splitting amplitude at three loops: 
\begin{align}
\nonumber
&\Delta_{\rm \bf Sp}^{(3)}=\left.(\Delta_n^{(3)}-\Delta_{n-1}^{(3)})\right\vert_{1\parallel 2}=-24\, (\zeta_5+2\zeta_2\zeta_3)\,\\
&\times\Big[f^{abe}f^{cde}\left\{{\rm \bf T}_1^a,  {\rm \bf T}_1^c\right\}\left\{{\rm \bf T}_2^b,  {\rm \bf T}_2^d\right\} +\frac12C_A^2  {\rm \bf T}_1\cdot {\rm \bf T}_2\Big] \,.
\label{eq:Delta_Sp_3}
\end{align}
We note that $\Delta_{\rm \bf Sp}^{(3)}$ only depends on the colour degrees of freedom of the collinear pair, and is entirely independent of the kinematics, and hence fully consistent with general expectations~\cite{SplittingAmplitudeRefs}\footnote{We recall that strict collinear factorization is restricted to time-like kinematics with both collinear partons in the final state, but it is violated for space-like splitting~\cite{Catani:2011st}.}. We emphasise that $\Delta_{\rm \bf Sp}^{(3)}$ is independent of the value of $n$ that was used to compute it, which is remarkable. 
Indeed, the fact that the difference in Eq.~\eqref{eq:Delta_Sp_3} is independent of $n$ requires intricate relations between different sets of diagrams and thus provides a highly non-trivial check of the calculation. 

To conclude, we have computed all connected graphs contributing to the soft anomalous dimension in multi-parton scattering and determined the first correction going beyond the dipole formula.
We find that such corrections appear at three-loops already for three coloured partons, but they only involve kinematic dependence in amplitudes with least four coloured partons, when conformally-invariant cross rations can be formed.
The final result is remarkably simple: it is expressed in terms of single-valued harmonic polylogarithms of uniform weight five.  Finally, we recover the expected behaviour of amplitudes in both the Regge limit and in two-particle collinear limits, and make further concrete predictions in both these limits.

\noindent
{\em Acknowledgments:} The authors are grateful to S. Caron-Huot for clarifying discussions on the role of colour conservation. We thank L. Dixon, G. Falcioni, M. Harley, L. Magnea, J. Pennington, J. Smillie, M. Schwartz, and C. White for useful discussions and collaboration on related topics. C.~D. acknowledges the hospitality of the Higgs Centre of the University of Edinburgh at various stages of this work. This research is supported by the STFC Consolidated Grants ``Particle Physics at the Tait Institute'' and ``Particle Physics at the Higgs Centre'' (E.~G.), by The University of Edinburgh via the PCDS PhD studentship and LHCPhenoNet (\O.~A.), as well as the ERC starting grant ``MathAm'' (C.~D.).


\begin{thebibliography}{99}

\bibitem{Mueller:1979ih}
A.~H. Mueller, 
  { Phys. Rev.} {\bf D20} (1979) 2037.

\bibitem{Collins:1980ih}
J.~C. Collins, 
{ Phys. Rev.} {\bf D22} (1980) 1478.

\bibitem{Sen:1981sd}
A.~Sen, 
{ Phys. Rev.} {\bf D24} (1981) 3281.

\bibitem{Sen:1982bt}
A.~Sen, 
{ Phys. Rev.} {\bf D28} (1983) 860.

\bibitem{Gatheral:1983cz}
J.~Gatheral, 
{ Phys.Lett.} {\bf B133} (1983) 90.

\bibitem{Frenkel:1984pz}
J.~Frenkel and J.~Taylor, 
 {  Nucl.Phys.} {\bf B246} (1984) 231.

\bibitem{Korchemsky:1985xj}
  G.~P.~Korchemsky and A.~V.~Radyushkin,
  Phys.\ Lett.\  B {\bf 171}, 459 (1986);
  G.~P.~Korchemsky and A.~V.~Radyushkin,
  Nucl.\ Phys.\ B {\bf 283}, 342 (1987);

\bibitem{Korchemsky:1993uz}
  G.~P.~Korchemsky and G.~Marchesini,
  Phys.\ Lett.\ B {\bf 313} (1993) 433.

\bibitem{Korchemsky:1993hr}
  G.~P.~Korchemsky,
  Phys.\ Lett.\  {\bf B325 } (1994)  459-466,
  hep-ph/9311294;
  I.~A.~Korchemskaya, G.~P.~Korchemsky,
  Phys.\ Lett.\  {\bf B387 } (1996)  346-354,
  [hep-ph/9607229];
  Nucl.\ Phys.\  {\bf B437 } (1995)  127-162,
  [hep-ph/9409446].

\bibitem{Magnea:1990zb}
L.~Magnea and G.~F. Sterman, 
{ Phys.Rev.} {\bf D42} (1990) 4222--4227.

\bibitem{Catani:1996vz}
  S.~Catani and M.~H.~Seymour,
  Nucl.\ Phys.\  B {\bf 485}, 291 (1997)
  [Erratum-ibid.\  B {\bf 510}, 503 (1998)], 
  [hep-ph/9605323].

\bibitem{Catasterm}
 S.~Catani,
 Phys.\ Lett.\ B {\bf 427} (1998) 161,
 [hep-ph/9802439];
 G.~F.~Sterman and M.~E.~Tejeda-Yeomans,
 Phys.\ Lett.\ B {\bf 552} (2003) 48,
 [hep-ph/0210130].

\bibitem{Dixon:2008gr}
  L.~J.~Dixon, L.~Magnea and G.~F.~Sterman,
  JHEP {\bf 0808} (2008) 022
  [arXiv:0805.3515 [hep-ph]].

\bibitem{Kidonakis:1998nf}
  N.~Kidonakis, G.~Oderda and G.~F.~Sterman,
  Nucl.\ Phys.\ B {\bf 531} (1998) 365
  [hep-ph/9803241].

\bibitem{Bonciani:2003nt} 
  R.~Bonciani, S.~Catani, M.~L.~Mangano and P.~Nason,
  Phys.\ Lett.\ B {\bf 575}, 268 (2003)
  [hep-ph/0307035].

\bibitem{Dokshitzer:2005ig}
  Y.~L.~Dokshitzer, G.~Marchesini,
  JHEP {\bf 0601}, 007 (2006),
  [hep-ph/0509078].

\bibitem{Aybat:2006mz}
  S.~M.~Aybat, L.~J.~Dixon, G.~F.~Sterman,
  Phys.\ Rev.\  {\bf D74 } (2006)  074004,
  [hep-ph/0607309].

\bibitem{Gardi:2009qi}
  E.~Gardi, L.~Magnea,
  JHEP {\bf 0903 } (2009)  079,
  [arXiv:0901.1091 [hep-ph]].

\bibitem{Becher:2009cu}
  T.~Becher, M.~Neubert,
  Phys.\ Rev.\ Lett.\  {\bf 102 } (2009)  162001,
  [arXiv:0901.0722 [hep-ph]];
  JHEP {\bf 0906 } (2009)  081,
  [arXiv:0903.1126 [hep-ph]].

\bibitem{Gardi:2009zv}
E.~Gardi and L.~Magnea, 
{  Nuovo Cim.} {\bf C32N5-6} (2009) 137--157,
  [arXiv:0908.3273 [hep-ph]].

\bibitem{Dixon:2009gx}
  L.~J.~Dixon,
  Phys.\ Rev.\ D {\bf 79} (2009) 091501
  [arXiv:0901.3414 [hep-ph]].

\bibitem{Dixon:2009ur}
  L.~J.~Dixon, E.~Gardi, L.~Magnea,
  JHEP {\bf 1002 } (2010)  081,
  [arXiv:0910.3653 [hep-ph]].

\bibitem{Bret:2011xm}
  V.~Del Duca, C.~Duhr, E.~Gardi, L.~Magnea and C.~D.~White,
  Phys.\ Rev.\ D {\bf 85} (2012) 071104
  [arXiv:1108.5947 [hep-ph]];
%
  V.~Del Duca, C.~Duhr, E.~Gardi, L.~Magnea and C.~D.~White,
  JHEP {\bf 1112} (2011) 021
  [arXiv:1109.3581 [hep-ph]].

\bibitem{Caron-Huot:2013fea}
  S.~Caron-Huot,
  JHEP {\bf 1505} (2015) 093
  [arXiv:1309.6521 [hep-th]].
  
\bibitem{Ahrens:2012qz}
  V.~Ahrens, M.~Neubert and L.~Vernazza,
  JHEP {\bf 1209} (2012) 138
  [arXiv:1208.4847 [hep-ph]].

\bibitem{Naculich:2013xa}
  S.~G.~Naculich, H.~Nastase and H.~J.~Schnitzer,
  JHEP {\bf 1304} (2013) 114
  [arXiv:1301.2234 [hep-th]].


\bibitem{Erdogan:2014gha}
  O.~Erdogan and G.~Sterman,
  Phys.\ Rev.\ D {\bf 91} (2015) 6,  065033
  [arXiv:1411.4588 [hep-ph]].

\bibitem{Gehrmann:2010ue}
T.~Gehrmann, E.~Glover, T.~Huber, N.~Ikizlerli, and C.~Studerus, 
  { JHEP} {\bf 1006} (2010) 094,
  [arXiv:1004.3653 [hep-ph]].






\bibitem{Kidonakis:2009ev}
N.~Kidonakis, 
  { Phys. Rev. Lett.} {\bf 102} (2009) 232003,
  [arXiv:0903.2561 [hep-ph]].

\bibitem{Mitov:2009sv}
A.~Mitov, G.~Sterman, and I.~Sung, 
{ Phys. Rev.} {\bf D79} (2009) 094015,
  [arXiv:0903.3241 [hep-ph]].

\bibitem{Becher:2009kw}
T.~Becher and M.~Neubert, 
{ Phys. Rev.} {\bf D79} (2009) 125004,
  [arXiv:0904.1021 [hep-ph]].

\bibitem{Beneke:2009rj}
M.~Beneke, P.~Falgari, and C.~Schwinn, 
{ Nucl.Phys.} {\bf B828} (2010) 69--101,
  [arXiv:0907.1443 [hep-ph]].

\bibitem{Czakon:2009zw}
M.~Czakon, A.~Mitov, and G.~F. Sterman, 
{ Phys.Rev.}  {\bf D80} (2009) 074017, [arXiv:0907.1790 [hep-ph]].

\bibitem{Ferroglia:2009ep}
A.~Ferroglia, M.~Neubert, B.~D. Pecjak, and L.~L. Yang, 
{  Phys.Rev.Lett.} {\bf 103} (2009) 201601,
  [arXiv:0907.4791 [hep-ph]].

\bibitem{Ferroglia:2009ii}
A.~Ferroglia, M.~Neubert, B.~D. Pecjak, and L.~L. Yang, 
   { JHEP} {\bf 0911} (2009) 062,
  [arXiv:0908.3676 [hep-ph]].

\bibitem{Chiu:2009mg}
J.-y. Chiu, A.~Fuhrer, R.~Kelley, and A.~V. Manohar, 
{ Phys.Rev.} {\bf D80} (2009) 094013,
  [arXiv:0909.0012 [hep-ph]].

\bibitem{Mitov:2010xw}
A.~Mitov, G.~F. Sterman, and I.~Sung, 
 { Phys.Rev.} {\bf D82} (2010)
  034020, [arXiv:1005.4646 [hep-ph]].

\bibitem{Gardi:2013saa}
  E.~Gardi,
  JHEP {\bf 1404} (2014) 044
  [arXiv:1310.5268 [hep-ph]];\,
  G.~Falcioni, E.~Gardi, M.~Harley, L.~Magnea and C.~D.~White,
  JHEP {\bf 1410} (2014) 10
  [arXiv:1407.3477 [hep-ph]].

\bibitem{Henn:2013tua}
  J.~M.~Henn, A.~V.~Smirnov and V.~A.~Smirnov,
  JHEP {\bf 1307} (2013) 128
  [arXiv:1306.2799 [hep-th]].

\bibitem{Gardi:2010rn}
E.~Gardi, E.~Laenen, G.~Stavenga, and C.~D. White, 
 { JHEP} {\bf 1011} (2010) 155,
  [arXiv:1008.0098 [hep-ph]].

\bibitem{Mitov:2010rp}
A.~Mitov, G.~Sterman, and I.~Sung, 
{ Phys.Rev.} {\bf D82} (2010) 096010,
  [arXiv:1008.0099 [hep-ph]].

\bibitem{Gardi:2011wa}
E.~Gardi and C.~D. White, 
 { JHEP} {\bf 1103} (2011) 079,
  [arXiv:1102.0756 [hep-ph]].

\bibitem{Gardi:2011yz}
E.~Gardi, J.~M. Smillie, and C.~D. White, 
{ JHEP} {\bf 1109} (2011) 114,
  [arXiv:1108.1357 [hep-ph]].



\bibitem{Gardi:2013ita}
E.~Gardi, J.~M. Smillie, and C.~D. White, 
{ JHEP} {\bf 1306} (2013) 088,
  [arXiv:1304.7040 [hep-ph]].


\bibitem{Sterman:1995fz}
  G.~F.~Sterman,
  In *Boulder 1995, QCD and beyond* 327-406
  [hep-ph/9606312].

\bibitem{Moch:2004pa}
  S.~Moch, J.~A.~M.~Vermaseren and A.~Vogt,
  Nucl.\ Phys.\ B {\bf 688} (2004) 101
  [hep-ph/0403192].

\bibitem{Grozin:2014hna}
  A.~Grozin, J.~M.~Henn, G.~P.~Korchemsky and P.~Marquard,
  Phys.\ Rev.\ Lett.\  {\bf 114} (2015) 6,  062006
  [arXiv:1409.0023 [hep-ph]].
 

\bibitem{Moch:2005tm}
S.~Moch, J.~Vermaseren, and A.~Vogt, 
{ Phys.Lett.} {\bf B625} (2005) 245--252,
  [hep-ph/0508055 [hep-ph]].
  
\bibitem{longpaper}
\O. Almelid, C. Duhr, E. Gardi, in preparation.

\bibitem{Czakon:2005rk}
  C.~Anastasiou and A.~Daleo,
  ``Numerical evaluation of loop integrals,''
  JHEP {\bf 0610} (2006) 031
  [hep-ph/0511176];\,
  M.~Czakon,
  ``Automatized analytic continuation of Mellin-Barnes integrals,''
  Comput.\ Phys.\ Commun.\  {\bf 175} (2006) 559
  [hep-ph/0511200];\,
  %
  A.~V.~Smirnov and V.~A.~Smirnov,
  ``On the Resolution of Singularities of Multiple Mellin-Barnes Integrals,''
  Eur.\ Phys.\ J.\ C {\bf 62} (2009) 445
  [arXiv:0901.0386 [hep-ph]];
  %
 {D.~A.~Kosower}, {\tt barnesroutines}, {\tt http://projects.hepforge.org/mbtools/}.

\bibitem{Anastasiou:2013srw}
  C.~Anastasiou, C.~Duhr, F.~Dulat and B.~Mistlberger,
  JHEP {\bf 1307} (2013) 003
  [arXiv:1302.4379 [hep-ph]].
  %

\bibitem{Goncharov.A.B.:2009tja}
A.~Goncharov, 
[arXiv:0908.2238 [math.AG]];
A.~B. Goncharov, M.~Spradlin, C.~Vergu, and A.~Volovich, 
{ Phys.Rev.Lett.} {\bf
  105} (2010) 151605, [arXiv:1006.5703 [hep-th]];
%
C.~Duhr, H.~Gangl, and J.~R. Rhodes, 
{ JHEP} {\bf 1210} (2012) 075,
  [arXiv:1110.0458 [math-ph]];
%
C.~Duhr, 
{ JHEP} {\bf 1208} (2012) 043,
  [arXiv:1203.0454 [hep-ph]];
  F.~C.~S.~Brown,
  Annales Sci.\ Ecole Norm.\ Sup.\  {\bf 42} (2009) 371
  [arXiv:math/0606419 [math.AG]];
  %
  F.~Brown,
  [arXiv:1102.1310 [math.NT]];
  %
  F.~Brown,
  Commun.\ Math.\ Phys.\  {\bf 287} (2009) 925
  [arXiv:0804.1660 [math.AG]];
  %
  J.~Ablinger, J.~Bl\"umlein, C.~Raab, C.~Schneider and F.~Wi\ss{}brock,
  Nucl.\ Phys.\ B {\bf 885} (2014) 409
  [arXiv:1403.1137 [hep-ph]];
%
  E.~Panzer,
  Comput.\ Phys.\ Commun.\  {\bf 188} (2014) 148
  [arXiv:1403.3385 [hep-th]].
  
\bibitem{BrownSVHPLs}
F.~C.~S.~Brown, 
C. R. Acad. Sci. Paris, Ser. I 338 (2004) 527.


\bibitem{Remiddi:1999ew}
  E.~Remiddi and J.~A.~M.~Vermaseren,
  Int.\ J.\ Mod.\ Phys.\ A {\bf 15} (2000) 725
  [hep-ph/9905237].
  

  
\bibitem{Dixon:2012yy}
  L.~J.~Dixon, C.~Duhr and J.~Pennington,
  JHEP {\bf 1210} (2012) 074
  [arXiv:1207.0186 [hep-th]].
  


\bibitem{Gaiotto:2011dt}
  D.~Gaiotto, J.~Maldacena, A.~Sever and P.~Vieira,
  JHEP {\bf 1112} (2011) 011
  [arXiv:1102.0062 [hep-th]].
  
\bibitem{Chavez:2012kn}
  F.~Chavez and C.~Duhr,
  JHEP {\bf 1211} (2012) 114
  [arXiv:1209.2722 [hep-ph]].

  
\bibitem{Smirnov:2003vi}
  V.~A.~Smirnov,
  Phys.\ Lett.\ B {\bf 567} (2003) 193
  [hep-ph/0305142].
  
\bibitem{Bern:2005iz}
  Z.~Bern, L.~J.~Dixon and V.~A.~Smirnov,
  Phys.\ Rev.\ D {\bf 72} (2005) 085001
  [hep-th/0505205].
  
  
\bibitem{Caron-Huot:TBA}
  S.~Caron-Huot, E.~Gardi, L.~Vernazza, ``Amplitudes in the high-energy limit,'' to appear.

\bibitem{SplittingAmplitudeRefs}
 Z.~Bern, V.~Del Duca, W.~B.~Kilgore and C.~R.~Schmidt,
 leading order,''
 Phys.\ Rev.\ D {\bf 60} (1999) 116001
 [hep-ph/9903516];\,
  D.~A.~Kosower,
  Nucl.\ Phys.\ B {\bf 552} (1999) 319
  [hep-ph/9901201];\, 
  I.~Feige and M.~D.~Schwartz,
  Phys.\ Rev.\ D {\bf 90} (2014) no.10,  105020
  [arXiv:1403.6472 [hep-ph]];\,
 S.~Catani, D.~de Florian and G.~Rodrigo,
 Phys.\ Lett.\ B {\bf 586} (2004) 323
 [hep-ph/0312067].


\bibitem{Catani:2011st}
 S.~Catani, D.~de Florian and G.~Rodrigo,
 JHEP {\bf 1207} (2012) 026
 [arXiv:1112.4405 [hep-ph]].
  

\end{thebibliography}
\end{document}